# Multi-Material 3-D Viscoelastic Model of a Transtibial Residuum from *In-vivo* Indentation and MRI Data

*David M. Sengeh, Kevin M. Moerman, Arthur Petron, Hugh Herr*\*

Center for Extreme Bionics, Massachusetts Institute of Technology, Cambridge, MA, USA

# Abstract

Although the socket is critical in a prosthetic system for a person with limb amputation, the methods of its design are largely artisanal. A roadblock for a repeatable and quantitative socket design process is the lack of predictive and patient specific biomechanical models of the residuum. This study presents the evaluation of such a model using a combined experimental-numerical approach. The model geometry and tissue boundaries are derived from magnetic resonance imaging (MRI). The soft tissue non-linear elastic and viscoelastic mechanical behavior was evaluated using inverse finite element analysis (FEA) of *in-vivo* indentation experiments. A custom designed robotic *in-vivo* indentation system was used to provide a rich experimental data set of force versus time at 18 sites across a limb. During FEA, the tissues were represented by two layers, namely the skin-adipose layer and an underlying muscle-soft tissue complex. The non-linear elastic behavior was modeled using 2nd order Ogden hyperelastic formulations, and viscoelasticity was modeled using the quasi-linear theory of viscoelasticity. To determine the material parameters for each tissue, an inverse FEA based optimization routine was used that minimizes the combined mean of the squared force differences between the numerical and experimental force-time curves for indentations at 4 distinct anatomical regions on the residuum. The optimization provided the following material parameters for the skin-adipose layer: $[c = 5.22\text{ kPa} \quad m = 4.79 \quad \gamma = 3.57\text{ MPa} \quad \tau = 0.32s]$ and for the muscle-soft tissue complex $[c = 5.20\text{ kPa} \quad m = 4.78 \quad \gamma = 3.47\text{ MPa} \quad \tau = 0.34s]$. These parameters were evaluated to predict the force-time curves for the remaining 14 anatomical locations. The mean percentage error (mean absolute error/ maximum experimental force) for these predictions was $7 \pm 3\%$. The mean percentage error at the 4 sites used for the optimization was 4%.

\*Corresponding author address: MIT Media Lab, 75 Amherst Street, Cambridge, MA 02139, USA. Telephone: +16173143661, email: hherr@media.mit.edu

# 1. INTRODUCTION

The prosthetic socket remains an important product in the life of a patient living with an amputation, yet its design and manufacture is still largely artisanal. As such, the production process is non-standard, non-repeatable and socket performance varies between manufacturers [1,2]. Therefore, many patients experience discomfort with their sockets due to improper fit, resulting in skin problems [3] including pressure sores, and deep tissue injury [4]. These skin issues are caused by the loading conditions associated with particular socket designs. Such loading conditions, or tissue stresses and strains, can be evaluated using computational modeling [5]. When combined with advanced computer-aided manufacturing techniques, computational modeling could also be a powerful tool for socket design optimization and biomechanical evaluation. Lee and Zhang (2007) presented a computational methodology for using pressure and pain evaluated on a residuum model to design better fitting sockets. While such a framework could enable prosthetists to design sockets in a more data-driven and repeatable manner, the authors assumed that the mechanical response of the soft tissue was linearly elastic with constants obtained from literature [6]. However, to enable computational design methodologies, finite element models of residuum should accurately describe the patient-specific geometry as well as the non-linear elastic and viscoelastic behavior of the underlying soft tissues.

This study focuses on the use of patient-specific *in-vivo* indentation and magnetic resonance imaging (MRI) combined with inverse finite element analysis (FEA) to determine the non-linear elastic and viscoelastic mechanical properties of an individual patient's residual limb. Such a FEA model is a stepping-stone towards quantitative socket design, as it would allow for the evaluation of loading conditions such as interface pressures and internal tissue stresses and strains. The residuum is, however, a complex multi-material structure consisting of the following main tissue types: skin, adipose, skeletal muscle, tendon and bone. Furthermore, the soft tissues undergo large non-linear deformations and are potentially subjected to high internal strains during prosthetic socket loading [5,7]. Finally, the geometry and biomechanical behavior varies from patient to patient, as such patient-specific analysis is required. Portnoy *et al.* (2009) concluded that patient-specific analyses of the residuum were important for evaluation of potential deep tissue injury from prosthetic devices [8]. Therefore, in order to ensure the fidelity of a residuum computational model, patient-specific analysis is required, multiple tissue regions need to be represented in the model, and the material behavior should capture the non-linear elastic and viscoelastic nature of the materials.

Previous soft tissue modeling research has been largely informed by animal tissue studies. Bosboom *et al.* (2001) presented an incompressible viscoelastic second-order

Ogden model that described skeletal muscle deformation. The elastic and viscoelastic properties were identified using a numerical-experimental procedure through invasive *in-vivo* compression tests on rat tibialis anterior muscles [9]. Van Loocke *et al.* (2008) performed compressive testing on porcine skeletal muscle samples demonstrating the anisotropic, non-linear elastic and non-linear viscoelastic behavior of skeletal muscle tissue [10,11]. The non-linear elastic and viscoelastic behavior were modeled using an extension of Hooke's law with strain-dependent Young's moduli, and a Prony series expansion, respectively. However, the elastic formulation used cannot easily be incorporated for computational modeling, and the parameters employed do not respect the constraints imposed by Hooke's law for transverse isotropy. To study soft tissue viscoelastic stress and shear response, Palevski *et al.* (2006) conducted a detailed study on porcine gluteus *in-vitro* and assumed muscle to be isotropic and linear elastic [12]. Although these animal studies offer an insight into the mechanical behavior of soft tissue, the results obtained cannot easily be translated to human applications let alone use for the residuum and socket design optimization.

The mechanical behavior of human tissues have been modeled and evaluated by other researchers. For example, to inform better micro needle designs, Groves *et al.* (2012) modeled a multilayer skin using 1st order Ogden material coefficients and evaluated it by using *in-vivo* indention experimental data [13]. Tran *et al.* (2007) used MRI and indentation to study the mechanical properties of human skin and muscle tissue modeled as a multi-layered neo-Hookean material [14]. The indentations in both studies were on the arm: the former applying small forces in comparison to loads on the residuum, while the latter used a two-dimensional model for analyses. Dubuis *et al.* (2011) used a mixed numerical-experimental method to study patient-specific soft tissue behavior of the lower limb through FEA compressive sock induced loading [15]. In that study, the adipose and skeletal muscle tissues were jointly modeled as a neo-Hookean material. The authors concluded that segmenting specific layers of the anatomy were useful for FEA approaches in order to understand internal tissue response.

While others have further used indenters to measure viscoelastic responses over various anatomical locations on human limbs, the conclusion was that the biomechanical material constants could not be readily extrapolated to other anatomical sites on the same residuum, or across separate residuum. Tönük and Silver-Thorn (2003) presented multiple reasons for the variability and lack of model predictability across the residuum. Their model simulations failed to converge at large deformations (>75% soft tissue thickness) and at thin but stiff regions [16]. Vannah and Childress (1996) also concluded that it was not possible to accurately and consistently model the biomechanical response of a bulk soft tissue across various locations on a limb using the same material constants [17]. Location dependent material constants make it difficult to integrate these models into quantitative socket design and other soft tissue modeling applications.

Recent work in residuum soft tissue modeling include studies describing the impact of socket design on internal soft tissues of the residuum [5,18], and those focused on the surface pressures [19,20] and stresses at the socket-residuum interface [7]. To evaluate internal soft tissue deformation in the muscle flap of the residual limb during static loading within a socket, Portnoy *et al.* (2008) used a computational model composed of two materials, the skin, and an internal soft tissue attached to rigid bones [5]. A neo-Hookean strain energy function described the instantaneous stress response of the muscle tissue coupled with a Prony Series expansion to capture viscoelasticity. The skin was modeled with a James-Green-Simpson strain energy function using material constants from literature (Hendriks *et al.* (2003) [21]). The residuum model was evaluated by comparing peak pressures measured with sensors within a custom cast/socket with those predicted by the combined residuum-cast model after the boundary conditions were applied. The peak pressures varied within 10 kPa between the experimental and simulation data. With all constitutive soft tissue material parameters obtained from literature rather than from patient-specific investigations, the authors limited their study by a lack of appropriate constitutive data. However, the conclusions about inhomogeneous internal compressive stress and strain distributions from that research especially around bony areas could be used to inform the design of quantitative prosthetic interfaces and further motivates the goal of developing predictive patient-specific validated residuum models.

The objective of this study is thus to advance a patient-specific, multi-material 3-D model of a transtibial residuum for a single patient, which would allow for the evaluation of loading conditions on the residuum from a prosthetic socket. We hypothesize that a FEA model composed of two layers of homogeneous materials (i.e. constant properties across the limb) can describe the non-linear elastic and viscoelastic tissue behavior for indentations across the limb. To evaluate this hypothesis, we used a combined experimental-numerical approach. A 3-D FEA model of a residual limb was created based on segmentation of detailed MRI data. Two tissue material were specifically modeled, a skin-adipose layer and an internal muscle-soft tissue complex. The parameters for these materials were then evaluated using inverse FEA based optimization to match the force boundary conditions from experimental indentation tests. A custom designed robotic *in-vivo* indentation system capable of loading the residuum at controlled rates is used to acquire a rich experimental data set of corresponding force versus time at 18 different anatomical locations across the residual limb. The tissue non-linear elastic material behavior was modeled by hyperelastic and 2nd order Ogden formulations while viscoelasticity was added through the quasi-linear theory of viscoelasticity. The experimental force versus time curves obtained for controlled load rates from the robotic indentation system are used as boundary conditions (load curves) for the inverse FEA based material parameter optimization. To determine the material parameters of the residual limb, an optimization routine is used that minimizes the difference (the combined mean of the squared force differentials) between the numerical and experimental force-time curves at 4 distinct anatomical regions on the residuum. The further evaluate the predictability of the FEA model, with optimized parameters for the two tissue layers, the experimental force-time curves for the remaining 14 anatomical locations were then predicted and compared to the experimental measurements. The predictive and patient specific model of the residuum presented, featuring material parameters evaluated based on in-vivo indentation, may prove critical to the future advancement of quantitative methodologies for prosthetic socket design.

## 2. METHODS

All data processing and visualization was performed using custom MATLAB (R2015a The Mathworks Inc., Natick, MA) codes and the open source MATLAB toolbox GIBBON (r110, [22,23], http://www.gibboncode.org/). FEA was implemented using the open source FEA software FEBio [24] (version 2.1.1, Musculoskeletal Research Laboratories, The University of Utah, USA, http://febio.org/).

### 2.1. Experimental methods

To accurately characterize the biomechanical behavior of the residuum through inverse FEA, three distinct processes were integrated. Firstly, surface and internal geometry data of the residuum were captured via non-invasive MRI of the residuum while MRI compatible skin markers were attached at 18 selected locations. These locations of the markers were informed by two main reasons: 1) specific locations of relevance in prosthetic socket design (for example, patellar tendon, fibula head, distal tibia, and posterior wall), and 2) anatomical variance: markers were placed on regions of large muscle thickness, bony regions, as well as medial and lateral points of interest all around the residuum. Surface segmentation of the MRI data provided the geometric input for FEA. Secondly, a custom indentation device was used to record force, time and displacement data for all locations corresponding to those highlighted by the MRI markers. Finally, non-linear elastic and viscoelastic material constants that defined the residuum were identified through inverse FEA based optimization using the boundary conditions derived from the experimental indentation. This section first discusses the MRI acquisitions followed by a description on the indentation experiments.

### 2.1.1. Magnetic resonance imaging (MRI)

For this study a patient with a bilateral transtibial amputation was recruited (male, age 50, amputation at age 17, weight 77 kg, activity level beyond K3). The amputation of the patient was for traumatic reasons. Informed consent was obtained using a protocol approved by the Committee on the Use of Humans as Experimental Subjects at the Massachusetts Institute of Technology. The patient was placed prone and feet-first inside a 3 Tesla MRI scanner (Siemens Magnetom Tim Trio 3T, Siemens Medical Systems, Erlangen, Germany). All imaging was performed with a RF body coil wrapped around the residuum without causing tissue deformations. An Ultra-short $T_E$ MRI (UTE-MRI) sequence (e.g. [25]) was used, ($T_R/T_E$=5.8/0.1, acquisition matrix 256x256, 256 slices, voxel size 1.18x1.15x1.00 mm) for image data acquisition.

The indentation experiment was conducted outside the MRI environment. Therefore, to highlight the desired indentation sites during imaging, 18 MRI compatible Beekley PinPoint® markers (Beekley Corporation, One Prestige Lane Bristol, CT 06010) were attached to the skin surface prior to imaging. These marker attachment sites were also denoted on the skin surface using body-safe eyeliner. Figure 1 illustrates marker locations on the actual skin surface of the volunteer, and on the skin surface reconstructed from the segmented MRI data. The surface models used in the optimization did not include marker shapes as these were only used to quantitatively identify marker locations.

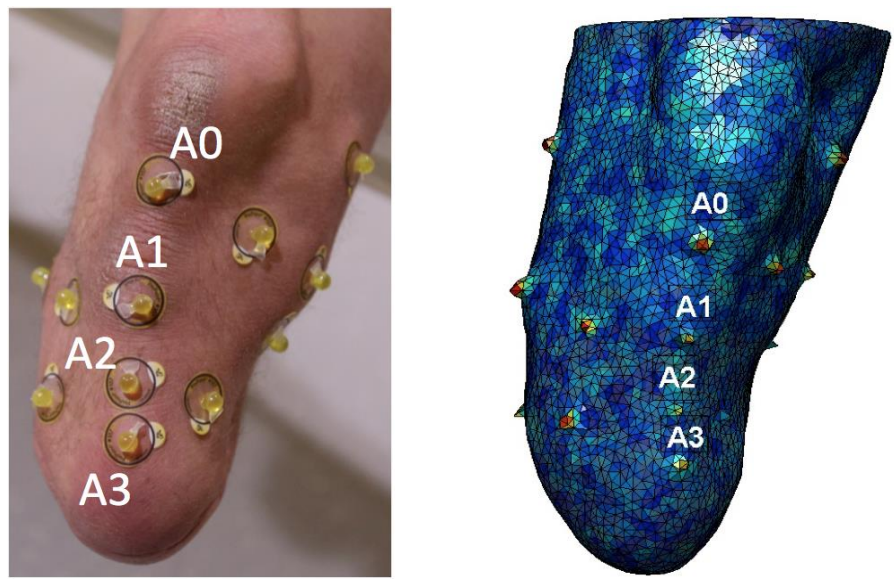

*Figure 1: MRI markers placed on the actual skin surface (left) and the corresponding marker regions highlighted on a surface model derived from the MRI data (right). A distance metric is used to quantitatively identify marker locations in the model (red locations corresponds to largest differences between the surface with a marker and the surface without, that is, marker locations)*

### 2.1.2. The indentation experiment

Immediately following MRI, indentation of the residuum was performed for all 18 sites using a custom designed and computer controlled indentation system named FitSocket (see also US Patent [26]). The FitSocket system is shown in Figure 2 and consists of a circular arrangement of 14 indentors. Each indentor head is a 20 x 20 mm non-rounded square block. The surfaces of the indentor heads are equipped with capacitive sensors allowing for detection of skin contact, i.e. moment of touch during loading.

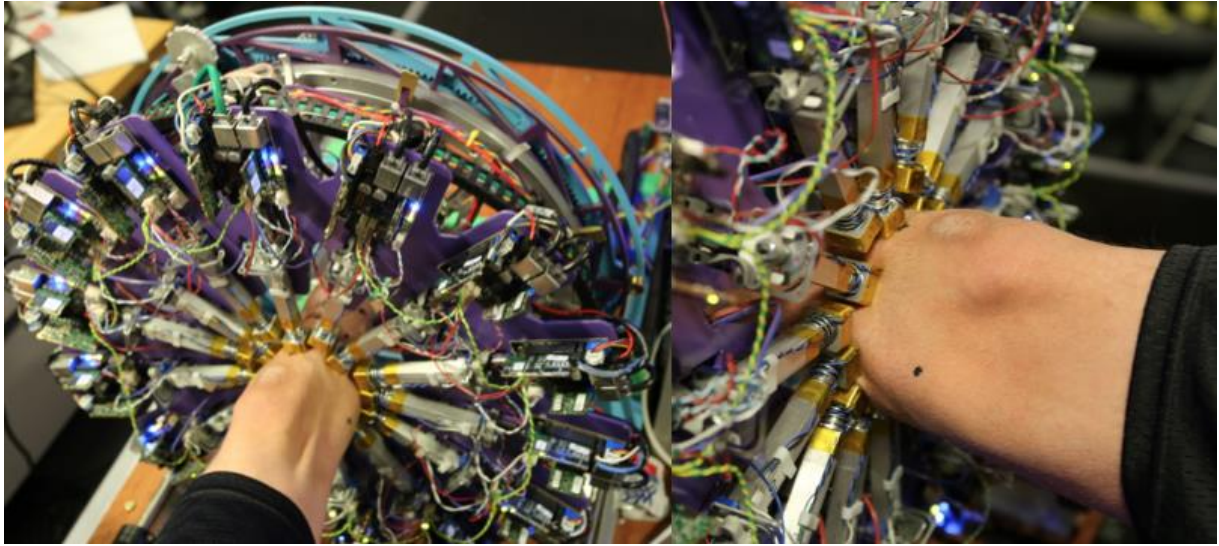

*Figure 2: An experimental setup showing a residuum within the FitSocket. Adjacent pins to the test pin (loading pin) are removed from the skin surface to allow tissue displacement*

The patient was seated comfortably next to the FitSocket system and asked to insert his residual limb into the device. The FitSocket system was then manually rotated and translated to position one of the indenters orthogonal to the skin at a test site. While the indentor positioned at the test site, called the test indentor, and its adjacent two indenters were held static (and were not touching the residuum) the other 11 indentors clamped the limb with an operator-selected force generally between 14 N and 16 N. The two adjacent indenters stayed removed from the skin surface to allow the tissue surrounding the indentation site to bulge during the indentation. Following clamping, all indentors were held in place while the test indentor was then activated to move towards its starting position to just touch the skin. This start position was determined by monitoring the indentor capacitance and force sensor data. Next, the maximum indentation depth was determined by slowly activating the indentor (at a rate of 5 mm/s) up to a maximum comfortable indentation level. This step allowed for the maximum achievable indentation depth to be set while patient discomfort was avoided. After recording this initial indentation used to set the maximum depth, the indentor was retracted to its initial starting position. A pause time of 5 s was then maintained to remove some pre-conditioning effects due to the initial test indentation. Then a single indentation was performed for the test site at a constant indentation speed 0.96 $\pm$0.5 mm/s. Although a constant indentation speed was used for all sites, local thickness variations meant that varying indentation depths and therefore strain rates were tested across the residuum. The experimental loading direction was such that during indentations, the flat surface of the indentor was always normal to the skin. As such, loading direction changed slightly during experimental indentation. During indentation, time, displacement and force were recorded at 500 Hz. Figure 3 shows a typical raw and regularized experimental force-time, force-displacement and displacement-time data. Regularization was performed to suppress the minor effects of noise. The regularized curves for displacement and force were derived from linear fitting, and cubic-smoothening spline fitting respectively. During regularization, the loading and unloading parts of the curve were treated separately (hence peak force is not smoothened). This initialization (orientation, alignment and maximum depth setting), indentation and regularization process was repeated for all 18 marked indentation sites.

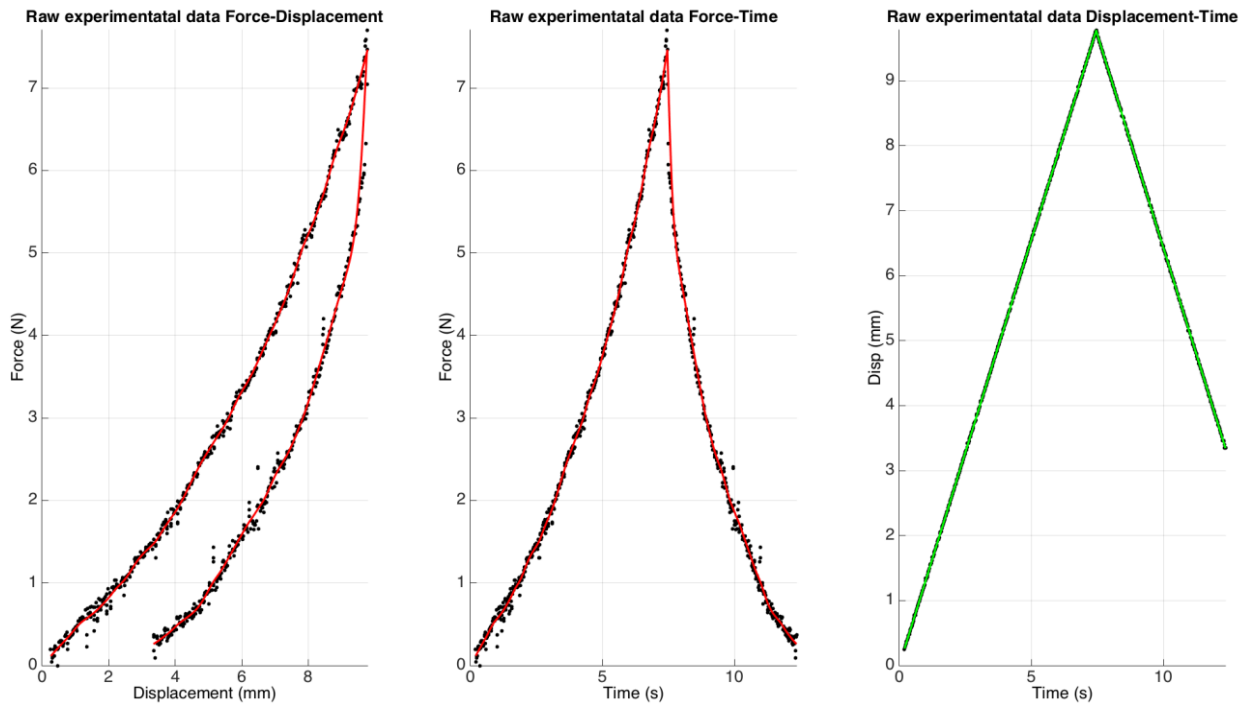

*Figure 3:* Typical raw FitSocket indentation experimental data. From left to right displacement-force, time-force and time-displacement curves are shown. Black dots denote the raw data while the solid curves are regularized curves.

## 2.2. Computational modeling

### 2.2.1. Finite element model construction: MRI segmentation → surface generation → meshing

For this study, tissue contours for the skin surface, muscle, and bones were segmented from the MRI data (based on GIBBON [22] *uiContourSegment* function). These contours were then converted to triangulated surface models. The two solid material regions modeled were: 1) skin (epidermis, dermis, and hypodermis/adipose layer), and 2) the remaining internal soft tissue (predominantly skeletal muscle and adipose tissue). The bones were represented as rigidly supported voids. The average thickness of the skin-adipose layer was observed to be 3 mm (consistent with thicknesses reported and used elsewhere [13,27–29]). Therefore, for simplicity the skin region was created with a homogeneous thickness of 3 mm by offsetting the outer skin surface inwards based on surface normal vectors. The solid material regions where meshed with 4-node tri-linear tetrahedral elements using TetGen (version 1.5.0, [www.tetgen.org](www.tetgen.org), see [30]) integrated within the GIBBON toolbox. The mesh density varied as a function of proximity to the indentation site with the smallest volume for elements close to the indentor and largest volume for those furthest away from the indentor. Mesh density was increased until the predicted indentation forces were no longer dependent on the mesh size.

For each of the 18 indentation sites, a dedicated FEA model mesh was constructed. At each site the central point of the flat head of the indentor was placed at the marker location derived from the MRI data. The indenter geometry, derived from its CAD design, was meshed using 5922 triangular shell elements and modeled as a rigid body. The indentor loading orientation orthogonal to the surface of the residuum was determined from the mean of the local skin surface normal directions. The indentor was then offset from the skin surface to avoid initial contact in the simulation. Figure 4 shows a typical segmented surface geometry and meshed 3-D FEA model geometry with the indentor model.

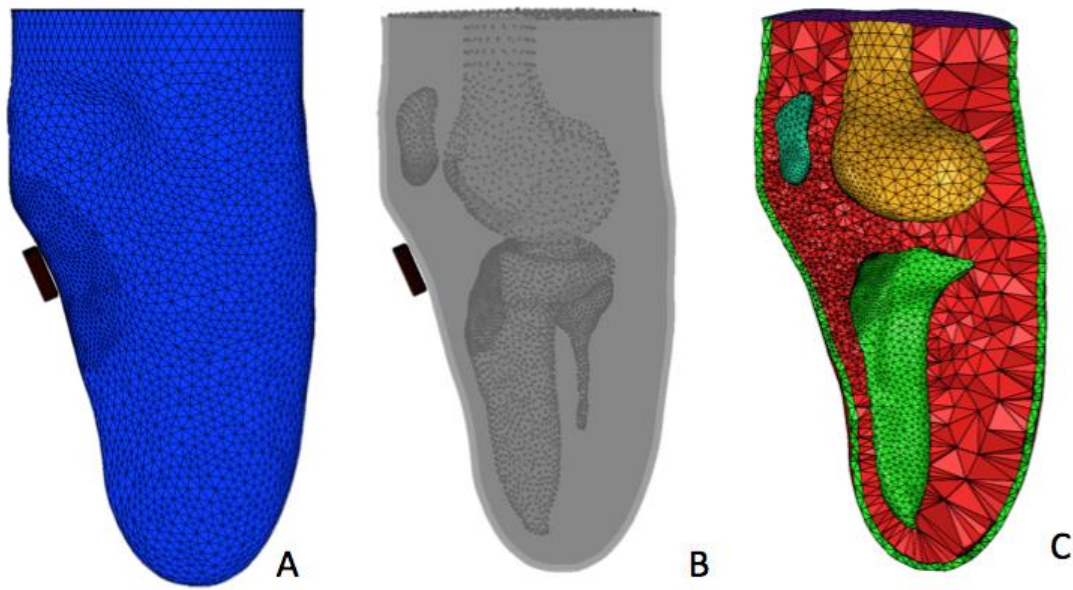

*Figure 4: (A) Typical surface model geometry showing local refinement near the indentor (example is for the patella tendon region), (B) transparent surface data showing supported internal surface nodes, (C) a typical solid tetrahedral mesh showing internal refinement as a function of proximity to the indentor. In addition, the two material regions, i.e. the skin-fat layer (green) and the internal soft tissue (red), and the bone voids are visible.*

## 2.2.2. Boundary conditions

The indentation boundary conditions (load curves for loading and unloading) of the simulation for each site were derived from the experimental displacement-time data. Therefore, indentation depth and rate of loading for each site corresponded to the experimental data. The slight variation in experimental loading direction was deemed little and thus it was assumed that the loading direction did not change in the simulation. A zero-friction sliding interface (see FEBio User Manual and also [31–34]) was assumed between the rigid indentor (master) and skin (slave) surface. All nodes of the top surface of the residuum and of the bones were constrained from moving in all directions (see Figure 4B). Hence the bones were represented by rigidly supported voids. Since the indentation sites were far from the top of the model, deformation in those regions were assumed to be negligible.

## 2.2.3. Constitutive modeling

The indentor was represented as a rigid body material. The soft tissue components were modeled as non-linear elastic and viscoelastic materials. Two soft tissue regions were distinguished: 1) a skin-adipose layer, and 2) an internal muscle-soft tissue complex. The patellar tendon was not separately modeled and was included in the internal soft tissue complex. The continuum mechanical formulations for these materials are briefly discussed below. For a detailed discussion of non-linear solid mechanics and tensor algebra the reader is referred to specialized literature [35–37].

The right Cauchy-Green tensor is given by:

$$\mathbf{C} = \mathbf{F}^\mathrm{T}\mathbf{F} \qquad\qquad 1$$

Where $\mathbf{F}$ is the deformation gradient tensor. The eigenvalues of $\mathbf{C}$ are the squared principal stretches $\lambda_i^2$. For FEA of nearly incompressible materials it is convenient to decompose deformation into deviatoric (isochoric and shape changing) and volumetric deformation. The following deviatoric deformation metrics can be defined:

$$\tilde{\mathbf{C}} = J^{-\frac{2}{3}}\mathbf{C}$$
$$\tilde{\lambda}_i = J^{-\frac{1}{3}}\lambda_i \qquad\qquad 2$$

With $J = \det(\mathbf{F})$ the volume ratio.

Elastic behavior

The elastic behavior is modeled using the following uncoupled, hyperelastic strain energy density function [38]:

$$\Psi = \frac{c}{m^2}\sum_{i=1}^{3}(\tilde{\lambda}_i^{\,m} + \tilde{\lambda}_i^{\,-m} - 2) + \frac{\kappa}{2}\ln(J)^2 \qquad\qquad 3$$

Here $c$, and $m$ are deviatoric material parameters, the former linearly scales the deviatoric response, while the latter controls the degree of non-linearity. This hyperelastic formulation is obtained from a second-order Ogden formulation with the parameters $c_1 = c_2 = c$ and $m_1 = -m_2 = m$ and has the tension-compression symmetry property $\Psi(\lambda_1, \lambda_2, \lambda_3) = \Psi\left(\frac{1}{\lambda_1}, \frac{1}{\lambda_2}, \frac{1}{\lambda_3}\right)$ (note this form reduces to a Mooney-Rivlin formulation if $J = 1$ and $m = 2$).

The volumetric behavior is dictated by the material bulk-modulus $\kappa$. The second Piola-Kirchoff stress tensor $\mathbf{S}$ can be derived from (see also [36,37]):

$$\mathbf{S} = 2\frac{\partial \Psi}{\partial \mathbf{C}} = 2\frac{\partial \widetilde{\Psi}}{\partial \mathbf{C}} + pJ\mathbf{C}^{-1} = J^{-\frac{2}{3}}\text{Dev}(\widetilde{\mathbf{S}}) + pJ\mathbf{C}^{-1} \qquad 4$$

With $p = \frac{\partial U}{\partial J}$, the hydrostatic pressure, and $\widetilde{\mathbf{S}} = 2\frac{\partial \widetilde{\psi}}{\partial \widetilde{\mathbf{C}}}$ is a deviatoric elastic stress. Use was made here of the deviatoric operator in the Lagrangian description:

$$\text{Dev}(\widetilde{\mathbf{S}}) = \widetilde{\mathbf{S}} - \frac{1}{3}(\widetilde{\mathbf{S}} : \mathbf{C})\mathbf{C}^{-1} \qquad 5$$

Given the high water content of biological soft tissue, near incompressible behavior is a common assumption. To achieve this, the bulk modulus is commonly set several orders of magnitude higher than the deviatoric stiffness parameters. During all simulations the bulk moduli were therefore constrained to be a factor 100 times higher than the elastic parameter $c$. This was found to be sufficient to enforce the volume ratio to remain within 1% of unity.

<u>Viscoelastic behavior</u>

Viscoelastic behavior is modeled using the quasi-linear theory of viscoelasticity (see also [39]). For the uncoupled formulations presented, the viscoelastic expression for the second Piola-Kirchoff stress can be written as [40]:

$$\mathbf{S}^v(t) = pJ\mathbf{C}^{-1} + J^{-\frac{2}{3}} \int_{-\infty}^{t} G(t-s)\frac{d\left(\text{Dev}(\widetilde{\mathbf{S}})\right)}{ds} ds \qquad 6$$

Here $G$ defines the following (single term) discrete relaxation function:

$$G(t) = 1 + \gamma e^{-t/\tau} \qquad 7$$

The parameters $\gamma$ and $\tau$ are proportional (units of stress) and temporal (units of time) viscoelastic coefficients respectively. It is clear that according to this formulation under static conditions eventually all viscoelastic enhancement can decay as a function of the viscoelastic parameters allowing equation 6 to reduce to the pure elastic stress defined by equation 4.

## 2.3. Inverse FE analysis based constitutive parameter optimization

This section describes the inverse FEA based constitutive parameter optimization. The iterative parameter optimization was done using custom MATLAB software capable of: 1) producing FEBio input files with the appropriate material parameters for the residuum-indenter model, 2) starting FEA analysis, 3) importing and analyzing the FEA results, 4) comparing FEA results to the experimental boundary conditions to formulate the objective function, and 5) performing inverse FEA based optimization of the objective function using a chosen optimization algorithm.

The inverse parameter identification employed Levenberg-Marquardt based optimization (implemented using the MATLAB *lsqnonlin* function, see also [41]).

The four indentation sites chosen for parameter optimization were representative of the different anatomical regions of a residuum used in socket design. The patella tendon region is used as a central point of reference in the design of conventional sockets. The tibia region is considerably different from the posterior wall region in geometry and material composition. The latter has a larger volume of soft tissue whereas the tibia region has little soft tissue between the skin surface and the bone. The final evaluation site was the lateral region between the tibia and the fibula. Anatomically, this region is between two bones and geometrically different from the other sites. The rest of the 14 locations distributed across the residuum were used to evaluate the model using the same material constants from the optimization.

The optimization was done in two stages (see Figure 5). Firstly, both tissue regions were treated as one leading to the optimization of four shared material parameters. A second optimization was then performed treating the two tissue regions as separate materials. The initial parameters for this second step were based on the optimal parameters of the first step. The optimization was deemed converged if either the parameters or the objective functions did not vary by more than 0.01.

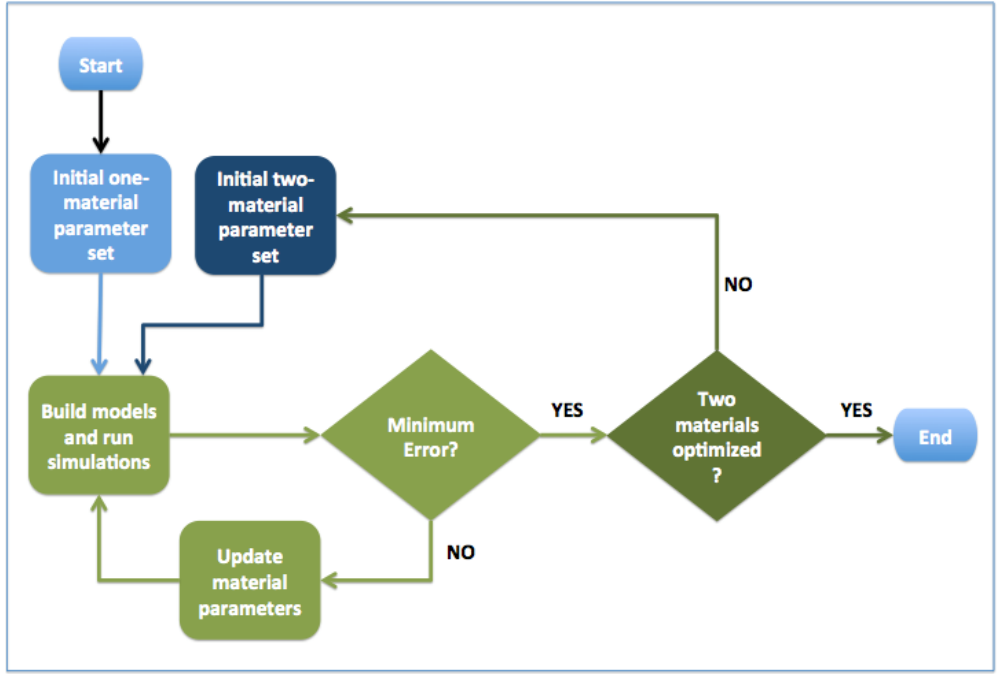

*Figure 5:* Schematic of two-step optimization routine used for the material constant identification.

The objective function vector $\mathcal{O}(\mathbf{p})$ was defined as:

$$\mathcal{O}_i(\mathbf{p}) = \frac{1}{n}\sum_{a=0}^{n}\left(F_{\exp_a}^{(i)} - F_{\sim_a}^{(i)}\right)^2 \qquad 8$$

Here $F_{\exp_a}^{(i)}$ and $F_{\sim_a}^{(i)}$ are experimental and simulated forces respectively, and $i$ and $a$ denote indentation site and time point indices respectively. Therefore $\mathcal{O}(\mathbf{p})$ is a vector whereby each entry reflects the squared differences of one of the four indentation sites. During the first step of the optimization procedure, a single material behavior is assumed leading to the material parameter vector $\mathbf{p}$:

$$\mathbf{p} = [c \quad m \quad \gamma \quad \tau] \qquad 9$$

After convergence of this initial step, the two material optimization employs the parameter vector:

$$\mathbf{p} = [c_s \quad m_s \quad \gamma_s \quad \tau_s \quad c_m \quad m_m \quad \gamma_m \quad \tau_m] \qquad 10$$

The subscripts s and m denote parameters belonging to the skin-adipose layer, and the muscle-soft tissue complex respectively. For the optimization, the parameter bounds were:

$$\text{minimum } \mathbf{p} = [c_s/100 \quad 2 \quad 0.01 \quad 0.01 \quad c_m/100 \quad 2 \quad 0.01 \quad 0.01]$$
$$\text{maximum } \mathbf{p} = [c_s * 10 \quad 20 \quad 10 \quad 10 \quad c_m * 10 \quad 20 \quad 10 \quad 10]$$

# 3. RESULTS

## 3.1. Dedicated and patient-specific FEA modeling of the residual limb

The residuum evaluated in this study was a patient-specific model with all easily distinguishable anatomical features including the surface of the skin and all the bones. Such an FE model could be used in the future to evaluate socket design and internal tissue deformations to understand the effects of surface loading on various anatomical features. In Figure 6, three different meshed models of a residuum are shown with bones (patella, tibia, femur are shown) represented as surface voids, a skin-adipose layer and an internal muscle-soft tissue volume meshed with tetrahedral elements. These models were for indentations at the patella region (left), the tibia region (center) and the posterior wall (right) and they show local mesh refinement for those regions. The residuum model can therefore be customized for different experiments.

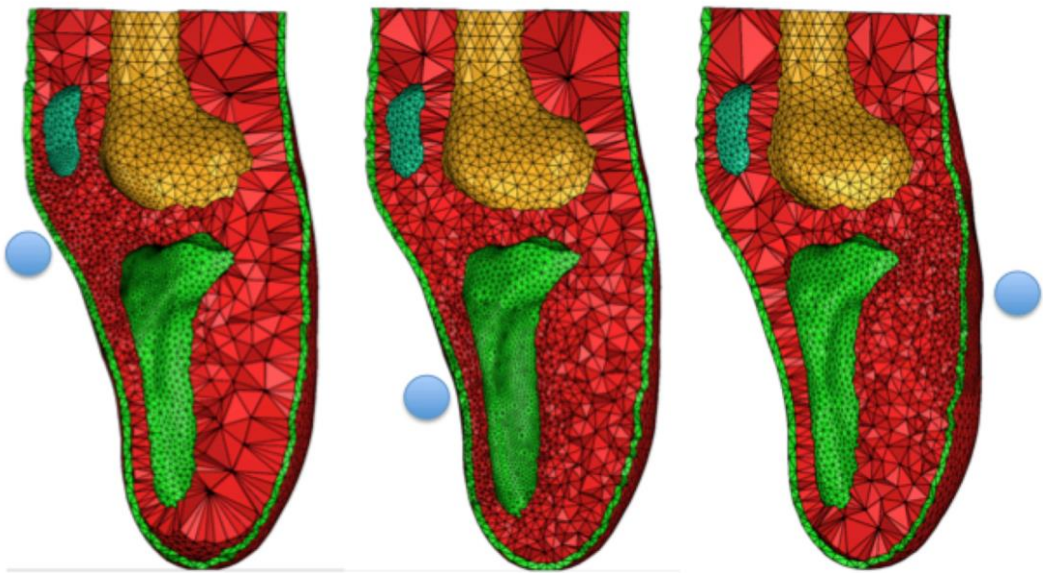

*Figure 6:* Three complete FEA models showing green elements as skin-adipose layer, bones as voids and red elements as internal soft tissue: patella tendon region (left), anterior tibia region (center), posterior wall (left). Blue markers show finely meshed regions

## 3.2. Inverse FEA based determination of the residuum constitutive parameters

It took between 10-60 minutes for an indentation simulation to converge in FEBio and to import the simulation results for analyses. The results of the optimization with material constants are summarized in Table 1. The optimization was done in two steps (see Figure 5) initially starting with a one material model and then using those optimum parameters as the initial input for the eight-parameter/two material residuum model. Those final optimized material constants presented below were then used to evaluate the mechanical response of the other 14 locations across the residuum.

| Tissue type | Material parameter | Initial (4 parameters) One material residuum model | Optimized value (4 parameters) One material residuum model | Optimized value (8 parameters) Two material residuum model |
|---|---|---|---|---|
| Skin-adipose layer | $C_s$ (kPa) | 4.7 | 5.2 | 5.22 |
| | $m_s$ | 3.00 | 4.74 | 4.79 |
| | $\gamma_s$ (MPa) | 1.20 | 3.86 | 3.57 |
| | $\tau_s$ (s) | 2.00 | 0.31 | 0.32 |
| Muscle -soft tissue complex | $C_m$ (kPa) | 4.7 | 5.2 | 5.20 |
| | $m_m$ | 3.00 | 4.74 | 4.78 |
| | $\gamma_m$ (MPa) | 1.20 | 3.86 | 3.47 |
| | $\tau_m$ (s) | 2.00 | 0.31 | 0.34 |

*Table 1: The initial and optimized constitutive parameters for a two-material transtibial residuum model*

Table 2 presents a summary of the maximum experimental loading force and the mean percentage error (average mean absolute error/ maximum experimental force) for all 18 indentation sites after the two step material optimization (the four indentation sites used in the optimization are denoted in blue).

| Loc | Max Experimental Force | 1 Material Residuum Model % Error | 2 Material Residuum Model % Error | Loc | Max Experimental Force | 1 Material Residuum Model % Error | 2 Material Residuum Model % Error |
|---|---|---|---|---|---|---|---|
| **1** | **8.3** | **6** | **6** | 10 | 12.31 | 6 | 6 |
| **2** | **8.1** | **4** | **4** | 11 | 11.46 | 11 | 11 |
| 3 | 8.8 | 6 | 6 | **12** | **10.33** | **4** | **4** |
| 4 | 15.5 | 7 | 6 | 13 | 7.46 | 1 | 1 |
| 5 | 14.4 | 2 | 2 | 14 | 7.91 | 9 | 9 |
| 6 | 13 | 5 | 5 | 15 | 10.78 | 6 | 5 |
| 7 | 16.7 | 13 | 13 | 16 | 10.89 | 4 | 4 |
| 8 | 12.7 | 5 | 5 | **17** | **12.21** | **4** | **4** |
| 9 | 12.3 | 11 | 11 | 18 | 11.51 | 10 | 10 |

*Table 2: Summary of results after optimization: locations in blue (1,2,12, and 17) were used in the optimization. (Loc=indentation location #)*

The force-time curves for the experimental and the simulation data for the four locations used in the optimization are presented in Figure 7. For the relatively stiff tibia region, the evaluated displacement is about 3.5 mm. This displacement is doubled at the posterior region, which is mostly soft tissue, and far away from bones.

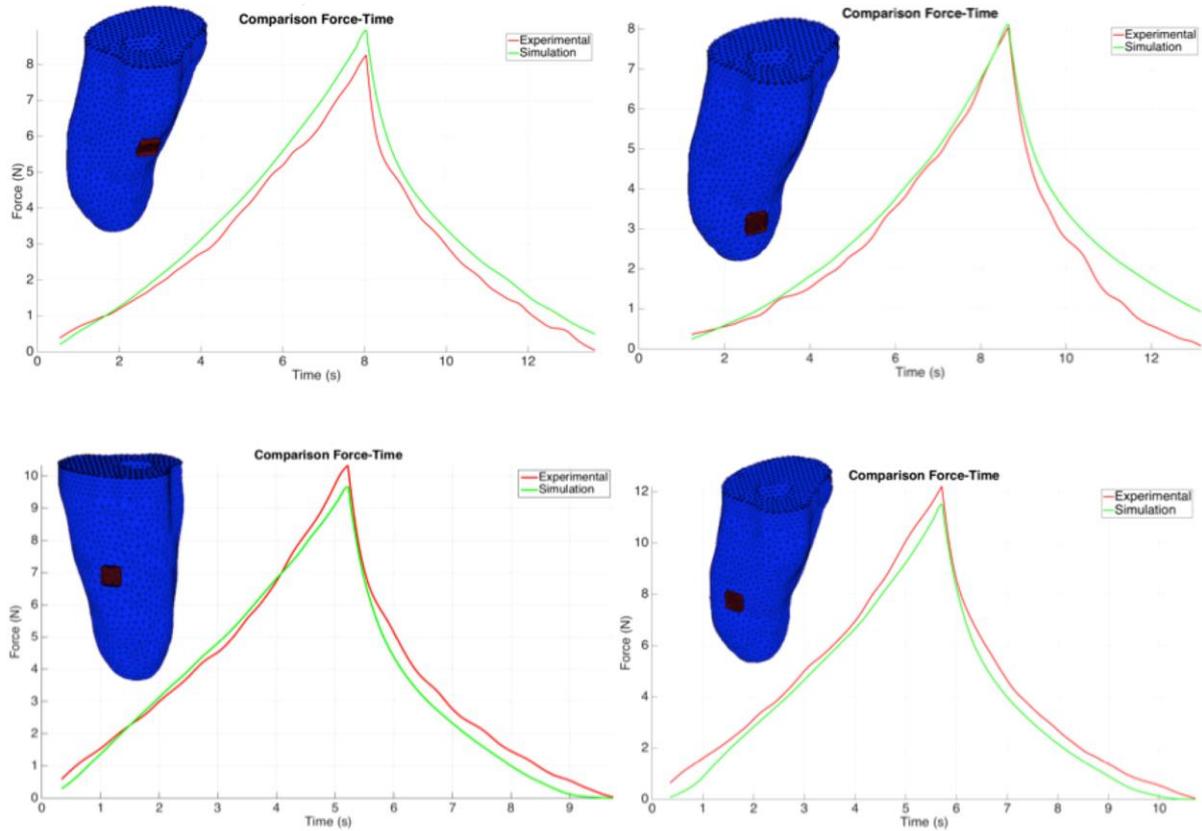

*Figure 7: Experimental and a simulated force-time curves at indentation site numbers 1, 2, (top row), 12 and 17 (bottom row) used in the optimization. These are the force-time curves using the material constants from the optimization.*

# 4. DISCUSSION

We hypothesized that a computational model composed of two layers of homogeneous materials (i.e. constant properties across the limb) can describe the non-linear elastic and viscoelastic tissue behavior at all anatomical points across the residuum of a person with a transtibial amputation. This paper consequently presented a combined experimental–numerical approach to define material constants for a two-material residuum model. The Ogden material parameters were derived from a non-linear optimization routine that minimized the combined squared differences of experimental and analytical force-time curves across four indentation sites of anatomical significance on the residuum.

The optimization was done in two steps. Firstly, the skin-adipose layer and the muscle-soft tissue complex were defined by the same parameters, and were thus effectively set as the same material. These initial optimized parameters were then used as the initial input from which parameters for a two material residuum model were derived. This staggered approach allows for the evaluation of both a single material residuum model and the investigation of a two material model. For this particular residuum, the mechanical response from the indentations is similar for a single material and a two material model. From these results, it can be concluded that a single bulk soft tissue volume could be used to effectively model the mechanical behavior of a residuum contrary to results reported by Tönük and Silver-Thorn (2003) [16]. While data was recorded *in-vivo* at multiple indentation sites across the residuum in our study and that of Tönük and Silver-Thorn, there are many differences in the approaches for instance in terms of geometry and material formulations used. Tönük and Silver-Thorn performed (non-linear) elastic simulations with 2-D axisymmetric models. In contrast, in the current study, we employed patient specific 3-D FEA and incorporated both non-linear elasticity and viscoelasticity.

Elastography techniques, e.g. based on MRI [42] and ultrasound [43] have also been used to estimate mechanical parameters of soft tissue of the lower limb. These studies present linear elastic shear moduli for muscle tissue of 3.73-7.53 kPa and 4.13 kPa (one third of the mean Young's Modulus) respectively. These are in reasonable agreement with the effective (initial) shear modulus derivable from the Ogden formulation employed here i.e. 5.2 kPa. However, elastography techniques assume Hooke's law of linear elasticity and as such do not capture large strain and non-linear hyperelastic behavior whereby elasticity is dependent on strain. Given the large strains [5] and pressures [20] seen during use of prosthetic sockets non-linear elastic behavior needs to be considered as presented here.

Our methodology uniquely combines: 1) non-invasive imaging, 2) patient-specific segmentation and FEA modeling, 3) a custom designed robotic and *in-vivo* indentation device, and 4) an inverse FEA based optimization of non-linear hyperelastic and viscoelastic material constants for various anatomical locations.

Comparison of the derived tissue material parameters to other studies is difficult due to the differences in methodology, tissue type, species of investigation, modeling approaches and constitutive formulations implemented. However, we will briefly discus other literature on soft tissue mechanical behavior. Van Loocke *et al.* (2008) described analysis of the transversely isotropic, non-linear elastic of excised porcine skeletal muscle tissue for *in vitro* compression using the strain dependent Young's moduli approach extended with Prony series to capture viscoelasticity [11]. Bosboom *et al.* (2001) used a first-order Ogden model to present a set of parameters that described the mechanical properties of skeletal muscle of rat under *in-vivo* compression ($c = 15.6 \pm 5.4 \text{ kPa}, m = 21.4 \pm 5.7, \gamma = 0.549 \pm 0.056 \text{ MPa}, \tau = 6.01 \pm 0.42 \text{ s}$) [9]. Lim *et al.* (2011) also presented material constants for a first-order Ogden model for pig skin (thickness of 2 mm) under dynamic tensile loading. Reported results for comparable strain rates were: $c = 20 \text{ kPa}, m = 11, c = 8 \text{ kPa}$ and $m = 7$ for loading parallel and perpendicular to the spine of the pig sample respectively [44].

To compare the material constants from these studies, a 10 x 10 x 10 mm cube described by the reported parameters was compressed for 0.5s, unloaded for 0.5s with an additional wait time at the end of 1 s and results evaluated. All boundary conditions and loading conditions were kept constant. Since Lim *et al.* did not have viscoelastic components, we added parameters from our research and not those suggested by Bosboom *et al.* based on the conclusion from Mukherjee *et al.* (2007) that their viscoelastic expansion was not ideal since loading and unloading paths were not the same [45]. Our material model for the human skin-adipose layer has a similar stress history curve to those predicted by Lim *et al.* when the constants for perpendicular loading were used as shown in Figure 8.

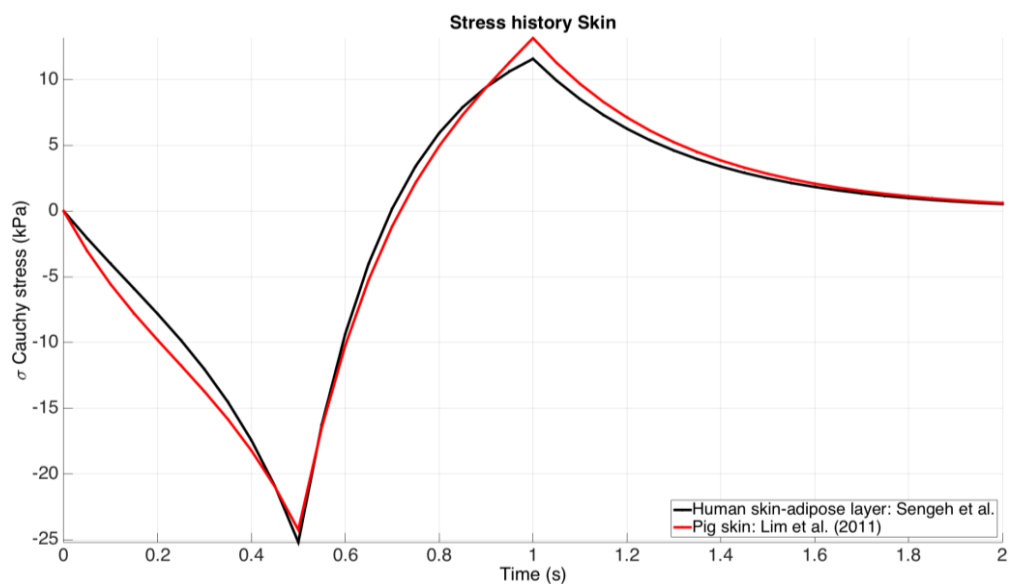

*Figure 8*: Comparing stress history of uniaxial compression for literature Ogden constants for skin

There is a noticeable difference in magnitude of stress and response decay observed between the rat tibialis muscle and the human muscle-soft tissue complex Figure 9. Perhaps this is expected since the data is not for human tissue and the composite/bulk response for soft tissue (adipose, tendons, skeletal muscle) is likely different from a skeletal muscle response. Further studies segmenting specific tissues and adipose would be necessary to get parameters for human skeletal muscle undergoing *in-vivo* loading.

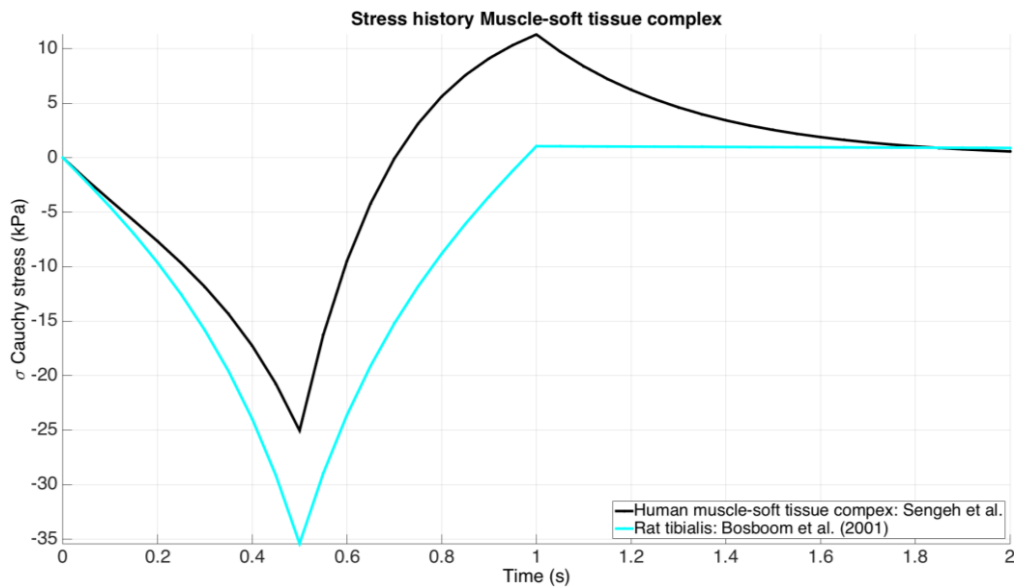

*Figure 9:* Comparing stress history of uniaxial compression for literature Ogden constants for muscle and muscle-soft tissue complex

The minimum and maximum errors between the force-time curves for the two material model simulation and the experimental setup across the limb were 2% and 13% respectively. The constitutive formulations used to describe the residuum model capture the elasticity, non-linearity and viscoelasticity observed in the residuum at all sites. There is little variation in results between the one material model and the two material residuum models. To better understand this further, more tissue segmentations should be investigated in the future. It must be noted that across the entire residuum, there are at least three regions of distinct biomechanical behavior: patella tendon region, hard body regions (along tibia for example) and soft body regions (in the posterior wall). An additional region would be the anterior medial and the anterior lateral regions, the latter assumed as a load bearing area for conventional socket design.

The 18 indentation sites were divided into these four regions. Each region constituting one of the locations used to characterize the entire residuum. Eight unique parameters for each region were derived as summarized in Table 3. The average regional errors when all points were separated into one of the four regions were 9%, 7%, 5% and 4% for region 1, 2, 3, and 4 respectively. Whether the parameters were tuned using individual regions or a combination of locations in all regions, the average error was consistent, across the residuum (7%). A sensitivity analysis of the *c* material constants for the skin-adipose layer and the muscle-internal soft tissue complex for an equal *m* showed that at thin regions of the body, the model was more sensitive to variations in the skin parameter. However, in thicker region like the posterior wall, the model was sensitive to changes in material constants for the muscle-soft tissue complex.

| Tissue type | Mat. param. | Initial material constants | Optimized across residuum Model (1, 3, 13, 20) | Optimized across 4 regions on residuum | | | |
|---|---|---|---|---|---|---|---|
| | | | | Region 1 **1**, 10, 11 | Region 2 **3**, 4, 8 | Region 3 **13**, 14, 15 | Region 4 5, 6, **20** |
| Skin-adipose layer | $c_s$ (kPa) | 4.7 | 5.22 | 5.4 | 5.5 | 5.1 | 5.2 |
| | $m_s$ | 3.00 | 4.79 | 9.86 | 6.27 | 4.68 | 7.55 |
| | $\gamma_s$ (MPa) | 1.20 | 3.57 | 2.73 | 2.64 | 3.59 | 1.70 |
| | $\tau_s$ s | 2.00 | 0.32 | 0.45 | 0.28 | 0.32 | 0.38 |
| Muscle-soft tissue complex | $c_m$ (kPa) | 4.7 | 5.20 | 5.1 | 4.0 | 5.2 | 5.5 |
| | $m_m$ | 3.00 | 4.78 | 4.52 | 5.21 | 4.75 | 3.85 |
| | $\gamma_m$ (MPa) | 1.20 | 3.47 | 2.49 | 2.66 | 3.59 | 3.04 |
| | $\tau_m$ s | 2.00 | 0.34 | 0.35 | 0.36 | 0.33 | 0.40 |

*Table 3:* Summary of results for optimum material parameters for each region compared to parameters across the residuum

In this study, a uniform skin-adipose thickness of 3 mm was assumed. This layer was a combination of the thin and stiffer epidermis and the thicker softer underlying adipose tissue. In the future, it would be worth segmenting the adipose layer from the skin layer particularly where the distribution of fat is not homogenous. The biomechanical behavior of skin and adipose tissue are very different and this may not be accurately reflected in the current combined form. Conclusions from Portnoy *et al.* (2006) indicate that presence of scar tissue, for example, can inform a more predictive patient-specific model [46]. Such patient-specific tissue features can be segmented from non-invasive image data and can be included in the computational modelling framework if deemed important.

Furthermore, it may be better for socket design if the patella tendon segmentation was included in the residuum model since most prosthetic socket designs rely on loading this tissue. Adding the stiffer patella tendon would potentially enhance the local force response for the computational model. However, at present, the match is already within 6% at this location. Tissue anisotropy should also be considered for future modeling. It also remains to be investigated whether muscle tone is an important feature and if local scarring or internal tissue adjustments may be relevant. As such active muscle modeling and spatially varying mechanical behavior can be included can be included in future work.

In defining the boundary conditions, a limitation and source of error was the direction of motion for the indentor. For the current indentation device, the indentor heads are not rigidly attached to their shafts but are instead able to alter their orientation somewhat during loading. This effect was not modeled and may have influenced the results for regions of high curvature where orientation changes may be expected like the fibula head region. In future experiments, the experimental loading direction must be quantitatively tracked using markers on the surface of the indenter and the residuum or a different rigidly attached indentor head needs to be installed. Other boundary conditions that affect convergence and the results of the simulation include contact conditions and the material bulk-modulus ($\kappa$). In this optimization $\kappa$ was set as 100 times the $c$ parameter, which was sufficient to enforce the volume ratio to remain within 1% of unity. With this value, there was convergence at all evaluated locations on the residuum.

The indentor geometry contains sharp edges and corners, which caused convergence difficulties for some simulations. In areas of high curvature on the residuum, it was more difficult to capture data because of the indenter shape and size. In this case, a spherical and smaller indenter would provide better data for loading around uneven surfaces. The indentor geometry also required a relatively high mesh density to allow the tissue to conform to these edges during indentation. For coarse meshes, penetration at these edges was observed, as the tissue mesh was unable to capture the edge geometry. As such to improve model convergence and potentially reduce mesh density (and therefore computational time), a smoother indentor geometry, such as a sphere, would be more desirable.

A zero-friction sliding interface assumption was used in this study. However, when a sticky contact was implemented, the maximum simulation forces did not vary significantly.

A further limitation in the evaluation presented here is the lack of validation of tissue deformation. Future work should incorporate the use of surface deformation measurement techniques, e.g. based on digital image correlation [47]. Alternatively, indentation experiments can be combined with simultaneous non-invasive imaging techniques such as MRI [48]. Such an approach would allow for the assessment of tissue geometry, and 3-D soft tissue deformation [49,50], and can also be combined with MRI based assessment of muscle fiber architecture [51]. This would allow for the detailed evaluation of the non-linear internal deformations as well as anisotropic material behavior. For future work, since the methodologies presented here are repeatable and use MRI data, we will model and evaluate other patient-specific residual limbs to better understand how these material constants vary across patients.

# 5. CONCLUSION

An important step in the process of quantitative prosthetic socket design is the development of a predictive biomechanical model of the residuum. This paper presents such a model for a single patient featuring non-linear elastic and viscoelastic constitutive behavior of residuum tissues. The model geometry was derived from non-invasive imaging and the constitutive parameters were evaluated based on *in-vivo* indentations. Although the inverse FEA optimization was based on only 4 distinct indentation sites on the residuum, the model was able to provide indentation force predictions for the remaining 14 sites on the residuum to within $7 \pm 3$ %.

# ACKNOWLEDGEMENTS

The Robert Wood Johnson Foundation (RWJF-ID 72293) and the MIT Media Lab Consortium funded this research. We would like to thank Steven Shannon, Atsushi Takahashi at the Martinos Imaging Center at MIT. In particular, thanks to Bryan Ranger in the Socket Team at the Biomechatronics Group. Many MIT undergraduates worked on this research and helped with MRI segmentation including Daivon Dean and Flora Liu.22